# Contraction response of a polyelectrolyte hydrogel under spatially nonuniform electric fields

E. Mert Bahçeci

*Processing and Performance of Materials, Department of Mechanical Engineering,
Eindhoven University of Technology, 5600 MB Eindhoven, The Netherlands*

Aykut Erbaş

*UNAM-National Nanotechnology Research Center and Institute of Materials
Science & Nanotechnology, Bilkent University, Ankara 06800, Turkey*

(Dated: August 20, 2024)

One of the major challenges in contemporary materials science is to build synthetic structures that can mimic responsive biological tissues, such as artificial skins and muscles. Polyelectrolyte hydrogels can provide such mechanoelectrical responses under external electric fields. Yet, their response to such stimuli, particularly at the molecular scales, is not fully revealed. Here, we study the mechanical response of a semi-infinite polyelectrolyte hydrogel slab to transient and spatially nonuniform sinusoidal electric fields by using extensive coarse-grained molecular dynamics simulations. Our simulations show that if the electric field is exerted on a small volumetric section of the hydrogel slab, the entire slab contracts reversibly and uniaxially in the direction perpendicular to the field. The hydrogel contracts almost half of its field-free initial length before retracting its original size, and eventually, its size fluctuations decay similarly to an underdamped oscillator. A contraction maximizes when the electric field is applied away from the boundaries of the hydrogel slab. Further, by tuning the electric field frequency and amplitude, both contraction times and efficiency can be controlled. Analyzes of contraction times and efficiency with implicit solvent simulations for various electrostatic parameters and salt concentrations confirm the robustness of the phenomenon while highlighting the importance of hydrodynamics. Our results demonstrate the effectiveness of spatially nonuniform electric fields to manipulate homogenous hydrogel structures for potential applications requiring rapid and soft actuation components.

## INTRODUCTION

An emerging component for biomimetic actuation systems is polyelectrolyte (PE) hydrogels [1–4], which are three-dimensional networks of crosslinked polymers with ionizable groups [5, 6]. Unlike conventional gels (e.g., rubber), PE hydrogels are highly solvent-swollen as a result of the balance between the entropic elasticity of constituting chains and osmotic pressure of ionized counterions entrapped inside the network structure [6–8]. This intricate balance gives PE hydrogels the ability to change their shapes in response to external stimuli, such as an electrical field, and bring them out as a novel component in actuator and ionotronics applications [9–14].

Despite the readily incorporation of hydrogels into electric-field-driven actuation systems, several challenges are standing in the way of designing rapid hydrogel-based actuation systems, particularly for applications requiring sophisticated deformation responses (e.g., for artificial muscles or skins). If a homogeneous hydrogel structure is placed between two electrodes and applied a voltage difference uniformly, hydrogel either shrinks isotropically or bends depending on the initial shape of the hydrogel [15, 16]. This is mainly due to the distinct effects of electrostatic forces on hydrogel components: while unconstrained counterions can move throughout the hydrogel structure towards the oppositely charged electrode under the electric field, PE chains are constrained by the network structure [17]. Consequently, resulting charge gradients across the hydrogel structure generate reversible but limited deformation responses [15, 18]. If the applied electric field has time-dependence (e.g., AC field), then the resulting deformation response could have a time-dependence as well.

To obtain more sophisticated shape changes, hydrogels are often combined with other materials, introduced structural heterogeneities, or pre-programmed [19]. For instance, when hydrogels are combined with dielectric elastomers as electrolytic elastomers, ionic double-layers at hydrogel interfaces lead to state-of-the-art rapid actuators [9, 10]. By introducing cross-linker gradients [20, 21], integrating multiple hydrogels with dissimilar chemistry [22–25], or by incorporating various structural nonuniformities [26, 27], a rich spectrum of deformation responses can be obtained. However, these methods inevitably bring additional complexities to fabrication steps, particularly as the hydrogel dimensions decrease, which is another critical requirement for fast response times [28–30].

Another way of obtaining shape deformations could be to apply a non-uniform electric field on a homogeneous hydrogel instead of increasing the complexity of the actuation system. Indeed, in most studies, electric fields operate uniformly on the entire hydrogel structure (i.e., dimensions of the electrodes are larger than those of the hydrogel), and thus, shape changes rely on structural or chemical nonuniformities of the hydrogel [2, 19]. Contrarily, if an electric field is applied to a small section of the hydrogel structure, field-free parts of the structure could be deformed via long-range interactions such as propagation of ionic gradients or polymer-network deformations throughout the structure. Deformation responses could be tuned by altering the frequency or amplitude of a time-dependent electric field or by decreasing the size of the hydrogel. In conjecture, a recent study demonstrated that spatially nonuniform multipolar electric fields could generate various 2D and 3D deformations on a homogeneous hydrogel sheet without any pre-programming or structural complexities [31]. To advance such applications, the response of hydrogels to nonuniform electric fields should be resolved in detail, which is an ideal task for computational calculations.

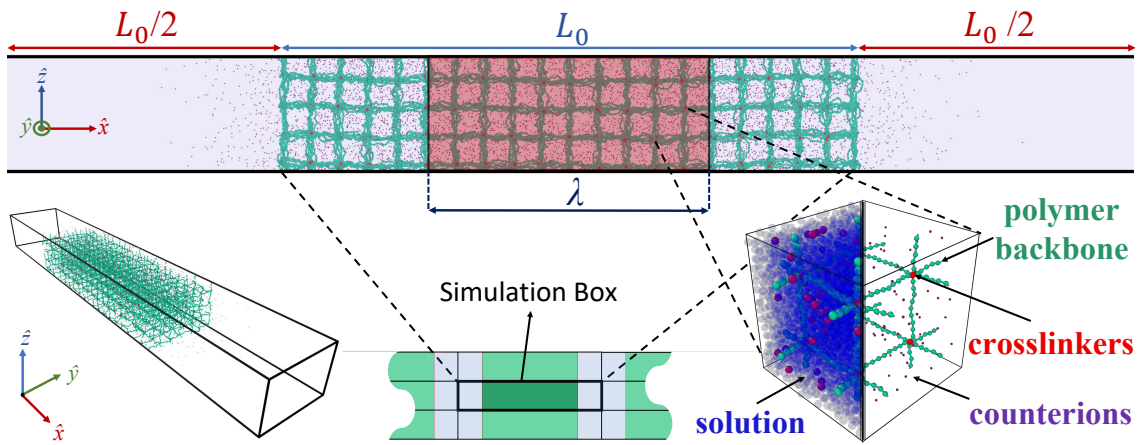

FIG. 1. The schematics of the MD simulations. The hydrogel slab used in the simulations is composed of defect-free cubic polymer lattices of $N$ monomers (green beads) with solvent beads (blue beads) and counterions (purple beads). The simulation box has a total width of $2L_0$ and is periodic in all directions. On both sides of the slab, there is a polymer-free void through which ions can diffuse in and out. The electric field is applied in the $\hat{z}$ direction only to a region of width $\lambda$ at the centre of the slab donated by a shaded rectangular area.

In this study, by using extensive coarse-grained molecular dynamics (MD) simulations of a semi-infinite homogeneous hydrogel slab (Figure 1) and considering both short and long-range electrostatics, we demonstrate that when a time-dependent sinusoidal electric field is applied to a small part of the hydrogel spatially nonuniformly, the slab contracts reversibly in the direction perpendicular to the applied field. The retraction of the hydrogel contraction follows an over-damped oscillatory pattern once the field is switched off. Excluding hydrodynamics by modelling the solvent explicitly preserves this response, albeit hydrogel oscillations change identity. Both the electrical field amplitude and frequency affect the contraction efficiency. The effect is more pronounced if the electric field is applied to the slab away from boundaries.

## METHODS

In MD simulations, PE chains of the hydrogels are constructed by using the coarse-grained Kremer-Grest (KG) bead-spring model together with their counterions and salt ions in explicit and implicit good-solvent conditions (Fig. 1) [8, 32, 33]. The hydrogel slab is constructed from defect-free cubic lattices composed of flexible chains (Fig. 1). The junction points of the lattices (i.e., crosslinkers) are permanent, and the number of beads between two crosslinkers is $N = 9$. To construct the slab, the unit lattice is repeated 20, 10, and 5 times in $\hat{x}$, $\hat{y}$, and $\hat{z}$ directions, respectively (Fig. 1). The slabs are finite in the $\hat{x}$-direction and periodic in the other two directions. In $\pm\hat{x}$ directions, there are hydrogel-free regions on each side between the hydrogels and the borders of the simulation box with dimensions $L_0/2$, where $L_0$ is the initial width of the slab (Fig. 1).

In the KG model, steric interactions between all the beads are calculated via a shifted 12/6 Lennard-Jones (LJ) potential (i.e., WCA),

$$U_{\mathrm{LJ}}(r) = \begin{cases} 4\epsilon[(\sigma/r)^{12} - (\sigma/r)^6 + 1/4] & \text{for } r < r_c, \\ 0 & \text{for } r > r_c, \end{cases}$$

where $\epsilon = 1 k_\mathrm{B} T$ is the energy unit, and $\sigma$ is the monomer unit size in the simulations. Here $k_\mathrm{B}$ is the Boltzmann constant, and $T$ is the absolute temperature. In Eq. 1, $r = |\mathbf{r}|$ is the distance between two beads, and $r_c = 2^{1/6}\sigma$ is the cut-off distance, which provides a repulsive LJ force between beads. Two adjacent chain beads, including crosslinkers, are bonded via a Finite Extensible Nonlinear Elastic (FENE) potential in the form of

$$U_{\mathrm{Bond}}(r) = -0.5 k r_0^2 \log[1 - (r^2/r_0^2)] \text{ for } r < r_0,$$

where $k = 50\epsilon/\sigma^2$ is the spring constant, and $r_0 = 1.5\sigma$ is the maximum bond stretching length. The Kuhn size of the flexible KG chains is $b \approx 1\sigma$.

A fraction of the polymer beads, $f = 0.5$, are assigned a positive unit charge $q = +1e$, where $e$ is the unit elementary charge. For each backbone charge, a counterion with a negative unit charge is added randomly in the simulation box to maintain the electro-neutrality condition (purple beads in Fig.1). The sizes of all ions and backbone monomers equal to $1\sigma$.

Short-range Coulomb interactions between all charges are calculated via

$$U_C(r) = -\epsilon \ell_B / r \text{ for } r < r_{ec},$$

where $\ell_B = e^2/(4\pi\epsilon_0\epsilon_s k_B T)$ is the Bjerrum length. Here $\epsilon_0$ and $\epsilon_s$ are the vacuum and solvent permittivity, respectively. The Bjerrum length quantifies the length scale, above which Coulomb energy between two charges is less than the thermal energy $1k_B T$ at $r = \ell_B$. The strength of the electrostatic interactions in the simulations is adjusted by tuning the dielectric constant of the medium to obtain $\ell_B \simeq 1\sigma, 2\sigma$. The case of $\ell_B \simeq 1b$ may correspond to water (i.e., $\ell_B \approx 7\text{Å}$). The case of $\ell_B > 1b$ may represent non-polar solvents at room temperature [34]. The electrostatic cut-off distance in Eq. 1 is $r_{ec} = 3\sigma$, above which longer-range electrostatic interactions are calculated via Particle-Particle-Particle Mesh (PPPM) Ewald solver. Various values for Ewald tolerances between $10^{-1}$ and $10^{-3}$ were considered, and no significant difference has been observed for the phenomenon reported here except for vanishing electric fields, at which thermal fluctuations are dominant (see Supplementary Information (SI) text).

Unless otherwise noted, the solvent is modelled explicitly in the simulations. To achieve this, hydrogel chains are relaxed in implicit solvent conditions (see below). Then, the simulation box is filled with beads of size $1\sigma$ to obtain a total particle volume fraction $\phi \equiv N_M \sigma^3/V \approx 0.8$, where $V$ is the volume of the entire simulation box and $N_M$ is the total number of beads in the box. Dissipative Particle Dynamics (DPD) for all pair interactions between the beads are employed for hydrodynamical effects [35]. In this method, the total force on a bead is divided into three contributions

$$\vec{f}_{dpd,i} = \sum_{j \neq i} \left( F_{ij}^C + F_{ij}^D + F_{ij}^R \right).$$

In Eq. 1, conservative $F^C$, dissipative $F^D$, and random $F^R$ forces are calculated as follows.

$$F_{ij}^C = \alpha \omega(r)$$

$$F_{ij}^C = -\gamma \omega^2(r)(\hat{r}_{ij} \cdot \vec{v}_{ij})$$

$$F_{ij}^R = \sigma \omega(r) \alpha (\Delta t)^{-1/2}$$

where $\alpha$ is a random number, $\sigma_{dpd} = \sqrt{2k_B T \gamma}$ with the thermostat parameter $\gamma = 1$, and the weight factor $\omega(r) = 1 - r/r_c$.

All MD simulations were run using LAMMPS MD package [36]. During the relaxation of solvent-filled initial configurations, a two-step procedure is followed: at the initial stage, each system is simulated for $10^6$ MD steps at a constant pressure $P = 0.05$ and $N_M$. The initial temperature is set to $T = 0.3\epsilon/k_B$ via Nosé-Hoover thermostat to avoid any numerical crash and gradually increased to $T = 1.0\epsilon/k_B$. The pressure-coupling constant of Nosé-Hoover thermostat is set to $\tau_p = 1.0\tau^{-1}$. For the second stage, the temperature is fixed to $T = 1.0\epsilon/k_B$, and the system is allowed to relax for another $10^7$ MD steps before the data-production runs are performed. Simulations are run with a time step of $\Delta t = 0.002\tau$. The unit time scale $\tau = \sigma\sqrt{m/\epsilon}$ is the LJ time step, whereas monomeric LJ mass $m$ is unity.

For a subset of simulations, the solvent beads are removed, and the solvent is modelled implicitly via Langevin dynamics, which is also the scheme used to relax the polymer network initially. In these simulations, the pressure is set to $P \approx 0.0$ in all directions to obtain a highly-swollen hydrogel slab [8, 12, 37].

A time-dependent sinusoidal electric field is applied only in the $\hat{z}$-direction on counterions and charged backbone beads within a prescribed section of the slab as shown in Fig. 1 via

$$\vec{E}(t) = E_0 \sin(\omega t)\hat{z}, \quad (1)$$

where $E_0$ is the strength of the electric field in the units of $\epsilon/q\sigma$, and $\omega$ is the frequency of the oscillation in the units of $1/\tau$.

Unless noted otherwise, results presented in this paper are averaged over time. Error bars are not shown if they are smaller than the size of the corresponding data point. Various numerical analyses on the MD trajectories are done using in-house analysis programs. See SI text for further benchmarking system sizes and electrostatic parameters.

**RESULTS**

**Electric field induces reversible oscillations on the PE hydrogel slab**

When an electric field is applied on a salt-free PE hydrogel, electrostatic forces move the backbone charges of PE chains and free counterions but in opposite directions. The counterions gain acceleration and continue their motion until they reach the boundaries of the hy-drogel (e.g., a non-penetrable interface). On the other hand, the motion of a backbone charge is constrained by the chain connectivity and permanent network topology. When the field is switched off, both chains and counterions return to equilibrium due to thermal fluctuations and intermolecular collisions. If the field is applied only on a small portion of the hydrogel slab, as shown in Fig. 1 by the blue shaded area, the question is how electric forces would affect the deformation of the rest of the hydrogel structure via propagation of chain deformations and ionic fluctuations.

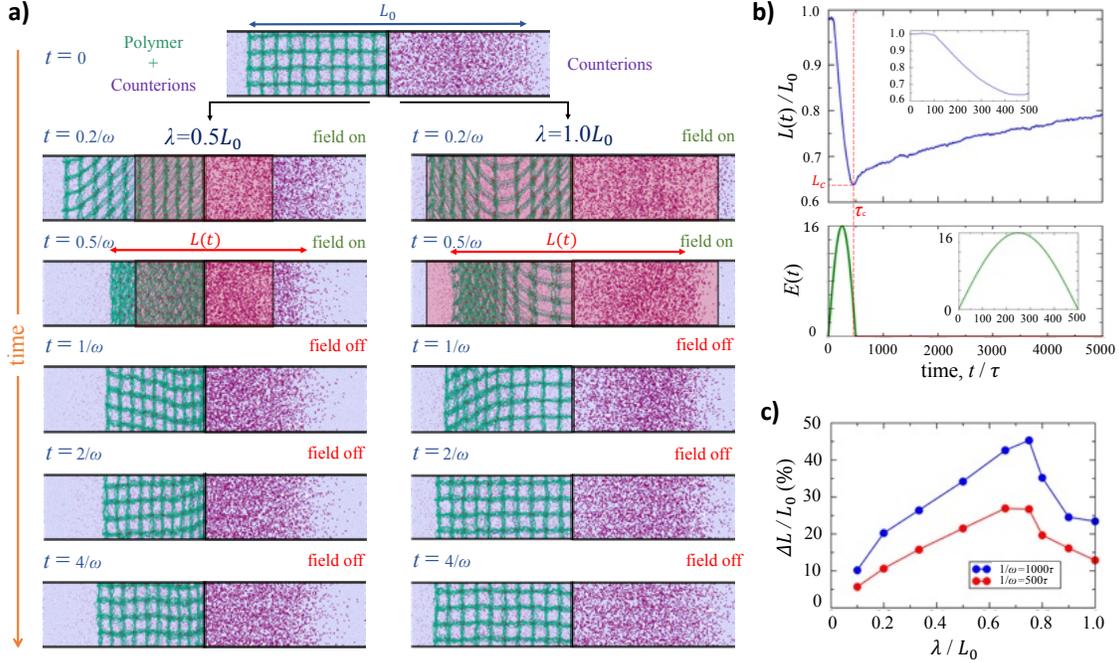

FIG. 2. a) Representative simulation snapshots showing the electric-field-driven contraction of a semi-infinite hydrogel slab. The electric-field pulse is applied for a time period of $1/500\tau$ on the central region of the slab indicated by a blue rectangle. The contraction and retraction of the slab are visualized at a fraction of the contraction time ($\tau_c$). For clarity, only counterions are visualized on the right side of the simulation box. b) The rescaled length of the hydrogel as a function of simulation time, $\tau$. The minimum contraction value is indicated by $L_c$, which defines the contraction time $\tau_c$. The inset shows the same data at shorter time scales. The bottom panel shows the sinusoidal electric field. c) The compaction efficiency of the slab, $\Delta L/L_0 \equiv (L_0 - L_c)/L_0$, as a function of $\lambda/L_0$ for two frequencies $\omega = 1/1000\tau^{-1}, 1/500\tau^{-1}$. The data points are joined to guide eyes.

Fig. 2a shows the representative simulation snapshots illustrating the time courses of deformation of the hydrogel slab under a sinusoidal electric field (see Eq. 1). The field is applied only to the central section of the slab with a width of $\lambda = L_0/2$ or $\lambda = 1.0 L_0$. We will refer to this central region where the field is applied as $\lambda$-region. The strength of the electric field is $E = 16\epsilon/q\sigma$. The signal is applied for a time period of $500\tau$ at a frequency of $\omega = 1/500\tau^{-1}$. During the mechanical response of the hydrogel to the applied electric field, we first observe the deformation of the polyelectrolyte chains in the field direction within the $\lambda$-region of the slab. The PE chains are stretched by the electrostatic forces (Fig. 2a). In the meantime, the distribution of counterions within the $\lambda$-region is also distorted (purple beads in Fig. 2a). Once the field is switched off, the chains and counterion clouds' deformation extends outside of the $\lambda$-region. Two deformation waves move towards both edges of the slab. The deformation fronts manifest themselves as denser density waves of polymer monomers (green beads in Fig. 2a) and counterions emanating from the centre of the hydrogel towards $\pm \hat{x}$ directions. Remarkably, as the deformation of the chains propagates across the hydrogel, the hydrogel slab contracts uniaxially in the direction perpendicular to the applied field (Fig. 2a). Once the hydrogel reaches its minimum size, a relaxation period starts, and the slab retracts to its initial dimensions (Fig. 2b). The deformed hydrogel slab relaxes similar to an over-damped harmonic oscillator, as shown (Fig. 2a,b).

As shown in Fig. 2a, the dimensions of the $\lambda$-region determine the amount of contraction. Fig. 2c, where the relative compaction of the hydrogel as a function of $\lambda$ normalized by initial hydrogel width is shown, demonstrates this behaviour quantitatively. If the field is applied on the entire hydrogel (i.e., $\lambda/L_0 = 1$), visually, a very weak contraction occurs (Fig. 2c). A similar weak deformation is also apparent if the field is applied less than 20% of the hydrogel. Nevertheless, the hydrogel reaches a maximum contraction at around $\lambda/L_0 \approx 0.7$, resulting in a non-monotonic dependence on $\lambda$. Above $\lambda/L_0 > 0.7$, a sharper decay is observed in the compaction response (Fig. 2c). We attribute this behaviour to the time-dependent increase of the PE chain density in the $\lambda$-region as the chains are dragged into the region. This effect becomes more dominant as $\lambda \to L_0$ (the right side of Fig. 2a). Nevertheless, the contraction efficiency systematically increases with $\lambda$ as long as the $\lambda$-region is sufficiently away from the hydrogel boundaries (Fig. 2b). Notably, deformation efficiency increases from 25% to 45% as the pulse frequency is 2-fold decreased (Fig. 2c). This suggests that if the PE chains in the $\lambda$-region are stretched for a longer time, the contraction increases. We will discuss this frequency-dependent contraction efficiency further in the following sections.

**Hydrogel initially contracts fast but relaxes slowly**

To further quantify the mechanical deformation response of the hydrogels, we calculate the contraction times, $\tau_c$, at which the hydrogel reaches its minimum length, $L_c$ upon the application of electric field (for a visual definition, see Fig. 2b). Fig. 3a shows the calculated values of $\tau_c$ as a function of the electric field strength for a frequency of $\omega = 1/500\tau^{-1}$. Within the electric field range spanning one order of magnitude, $\tau_c$ does not show a strong dependence on the field strength (Fig. 3a). Notably, $\lambda$ does not change this behaviour.

For the frequency dependence of $\tau_c$, we observe a monotonic increase with the inverse frequency such that at slow frequencies, $\tau_c$ is not affected by the frequency (Fig. 3b). Note that we report the data as a function of inverse frequency since this metric corresponds to the time the electric field is applied to the hydrogel. At fast frequencies, we observe a decrease in $\tau_c$ at $1/\omega^* \approx 10^3 \tau$. Interestingly, this threshold (inverse) frequency does not depend on $\lambda$ (Fig. 3b. One possibility is that the threshold values are related to the relaxation time of the entire slab. However, as shown in Fig. 2, the hydrogel retracts its original shape at time scales $t > 5000\tau$, considerably longer than the threshold time scale $1/\omega^*$.

We anticipate that the electric field is applied longer or shorter than the relaxation time of the chains between two crosslinkers can lead to the gradual increase observed in Fig. 4b. The relaxation time of an $N$-mer chain in a solution under hydrodynamic effects can be described by the Zimm model [38]. The Zimm relaxation time is defined as $\tau_Z \approx N^{3\nu}\tau_0$, where $\tau_0 \approx \tau$ is the relaxation time of a single monomer, and $\nu$ is the (inverse) fractal dimension of the chains. For $N \approx 10$ and $\nu = 1$ for stretched PE chains [39], the relaxation time is $\tau_Z \approx 1000\tau$, which is close to the threshold frequency $\omega^* \approx 1/\tau_Z$ observed in our simulations (Fig. 4b). Consistently, since the frequency is the inverse of time that all charged species, including PE chains, within the $\lambda$-region experience the electric forces, at the low frequencies (i.e., $\omega < \omega^*$), hydrogel chains are stretched slowly by preserving their equilibrium conformation. This, in turn, could result in a more efficient contraction of the hydrogel slab in the direction perpendicular to the field. As the frequency increases, the time is insufficient to overcome the drag forces to stretch and displace the chains strongly. Thus, contraction decreases.

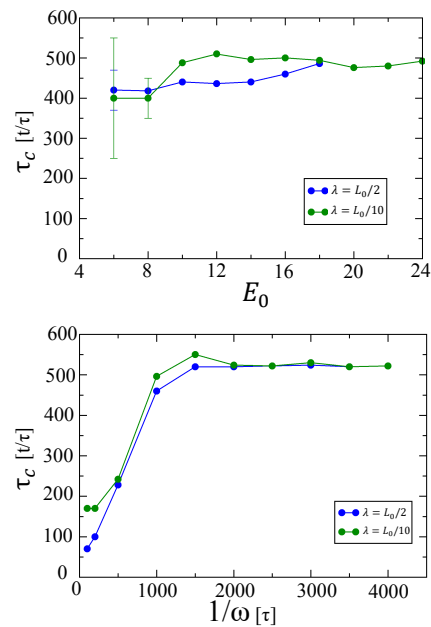

FIG. 3. Contraction time, $\tau_c$ as a function of electric field strength for $\lambda$ values. $\omega = 1/500\tau^{-1}$. The contraction time as a function of inverse frequency. $E_0 = 16\epsilon/q\sigma$. See Fig. 2 for the definition of contraction time.

We also calculate how quickly the entire hydrogel structure retracts its original shape. To do this, we fit the data beyond $t > \tau_c$ to an empirical single exponential function. Our analyses reveal that the relaxation of the slab is $\tau_R > 8000\tau$, which is much slower than the contraction time calculated (see SI text). Our data also indicate that the slab relaxation time is independent of $\lambda$ as expected since the entire hydrogel is deformed.

**Electric field strength and frequency can tune the contraction response of the hydrogel**

The field-driven compaction demonstrated in the previous section is a direct consequence of electrostatic forces acting on the charged backbone monomers and counterions inside the $\lambda$-region. This force acts for a time duration of $1/\omega$ for a pulse signal considered here. Thus, both electric field and frequency can determine the amount of compaction. In Fig. 4a, we show how the electric-field strength affects the contraction efficiency. Generally, the stronger the field is, the more the hydrogel slab contracts regardless of the frequency, and the effect of the field increases monotonically, as shown in Fig. 4a. For small $\lambda$ (e.g., $\lambda = L_0/10$), fewer deformed PE chains within $\lambda$-region lead to weaker contraction of the hydrogel. Increasing $\lambda$ gradually intensifies the effect of the electric field on the contraction efficiency. Note that the maximum electric field strength we can employ in our simulations is determined by the numerical bond errors, suggesting that covalent-bond ruptures would determine the upper limit of the field strength in experimental conditions.

The frequency of the electrical pulse also affects the contraction efficiency, albeit its effect is non-monotonic (Fig. 4b). As the frequency is increased, we observe an increase in the amount of perpendicular contraction. This increase reaches a maximum, depending on $\lambda$, and decreases to almost vanishing values at the maximum frequency we achieve here (i.e., $\omega = 10^{-3}10^{-3}\tau^{-1}$). Since the inverse frequency also determines the duration of exposure time of the hydrogel PE chains to the electric field, shorter exposure leads to weakly deformed chains in the $\lambda$ region at fast frequencies. Notably, the frequency at which we observe peak values for the compaction efficiency seems to be independent of $\lambda$ and occurs at $\omega^* \approx 10^{-3}\tau^{-1}$) (Fig. 4b). Remarkably, this value is consistent with the threshold frequency $\omega^* \approx 1/\tau_Z$ calculated in the previous section.

### Absence of hydrodynamics and explicit solvent changes the characteristics of shape relaxation but not the contraction response

The contraction response that we observe in our simulations requires an interplay between solvent filling the pervaded volume of the hydrogel and the deformation dynamics of the polymer network. As the hydrogel deforms under the electric field, solvent molecules (modelled by beads in the simulations) must evacuate the hydrogel volume, allowing the hydrogel to deswell (Fig. 2a). In this process, the solvent can also change the relaxation dynamics of the network chains via hydrodynamic interactions. To further assess the role of hydrodynamics and the explicit nature of the solvent, we run simulations with an implicit solvent model. In this model, all solvent molecules are removed from the simulation box, but the effect of solvent viscosity is taken care of by Langevin dynamics (see Methods section for details). While implicit solvent simulations can model the equilibrium swelling behaviour of hydrogels [7, 12], they cannot model the hydrodynamic interaction or any effect stemming from the steric interactions between solvent and polymer molecules.

In Fig. 5, we compare our explicit simulations with implicit solvent simulations for various electric field strengths and frequencies. One immediate result of the implicit solvent is the amplified effect of the electric field; under equal electric field and frequency, the hydrogel with implicit solvent contracts more than the hydrogel with explicit solvent (Fig. 5a,b). Electric field strengths on the order of $E_0 \simeq 1\epsilon/q\sigma$ that do not contract the hydrogel in the explicit solvent simulations can significantly deform the hydrogel (Fig. 5a). In fact, to obtain a similar contraction efficiency of around $\Delta L/L_0 \approx 30\%$, explicit solvent requires almost 4-fold stronger electric field strengths (Fig. 5a). We attribute this behaviour to the osmotic (solvent-induced) pressure built up inside the hydrogel as the slab contracts, which does not exist in the implicit simulations. The contraction becomes more efficient even under low electric field strengths in the implicit solvent simulations since there is no solvent to displace. A similar response pattern is also apparent for the frequency dependency; in implicit simulations, the hydrogel can shrink at much lower frequencies as compared to the explicit solvent simulations, albeit a monotonic behaviour is observed as the frequency is increased (see SI text).

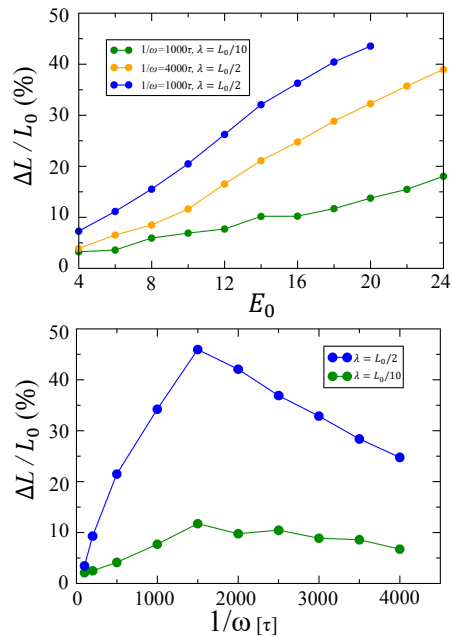

FIG. 4. Contraction efficiency, $\Delta L/L_0$, for various electric field strengths $E_0$, frequencies $\omega$, and electric field applied areas $\lambda$. a) The contraction efficiency as a function of electric field strength is scanned for two different frequency values $\omega = 1/1000\tau^{-1}, 1/4000\tau^{-1}$, and for two different electric field applied areas $\lambda = L_0/2, L_0/10$. b) The contraction efficiency as a function of frequency is scanned for two different electric field applied areas $\lambda = L_0/2, L_0/10$ for electric field amplitude of $E_0 = 16\epsilon/(q\sigma)$.

Once the field strength and frequency are adjusted, the initial contraction of the hydrogel with or without solvent is qualitatively similar (Fig. 5a,b). Nevertheless, the presence of solvent and, thus, hydrodynamics signif-icantly alter the relaxation dynamics of the hydrogel at most (Fig. 5). In our default case, the explicit solvent, hydrogel slab takes its original shape like an overdamped oscillator, but this situation switches to an under-damped oscillation in the implicit simulations (Fig. 5a,c). This does occur in the hydrogel swollen by explicit solvent molecules since evacuating these molecules from the hydrogel does not allow an over-damped oscillatory behaviour since this process is much slower. Overall, the comparison between implicit and explicit solvent simulations highlights the importance of hydrodynamics in the observed contraction response but also posits that it is the PE hydrogel network topology and its constrained backbone charges/counterions but not the solvent that results in the reversible contraction of the hydrogel in our simulations.

### Effect of excess salt on the contraction response

Another factor that can influence hydrogels' response to electrical fields is the excess salt. The salt ions screen the electrostatic (pairwise) interactions, thus reducing the repulsion/attraction between charges. This reduction can decrease the equilibrium swollen size of the PE hydrogels [40]. However, if the salt concentration exceeds a threshold limit, ion-ion correlations can cause a re-entrant transition that can swell the hydrogel [41]. Such shrinkage of the network structure can, in turn, influence the contraction response to the external electric field.

Having established that implicit and explicit solvents can result in qualitatively similar (initial) contraction responses (Fig. 5), for computational efficiency, we use the implicit solvent to demonstrate the effect of salt on the initial contraction of the hydrogel. This approximation reduces the computational time from several days to less than a day for our system size. In simulations, monovalent salt ions, also charged beads with unit sizes similar to the counterions, are added to the simulation box at random positions to obtain a prescribed salt concentration. The addition of salt up to a concentration of 10 mM results in a small effect on the swollen size of hydrogels (e.g., 27% shrinkage at 10 mM salt), but this difference is visually undetectable despite the apparent increase of the ionic concentration (Fig. 6a). Higher salt levels are not feasible even in the implicit solvent simulations due to the immense computational requirements to simulate such large systems by including all short and long-range electrostatics. Nevertheless, although the shrinkage of hydrogels by salt leads to weaker oscillations, size fluctuations are qualitatively similar to the salt-free cases (solid curve in Fig. 6b). Despite the agreement, as the salt levels increase, the contraction tends to weaken (Fig. 6b). Plotting the contraction as a function of salt concentrations also reveals this effect: as the salt concentration is increased, the hydrogels tend to contract less under the electric field as compared to their salt-free solutions (Fig. 6c). Thus, we conclude that the phenomena we observe can either decrease or diminish as the hydrogel becomes less swollen and/or electrostatic screening increases.

### CONCLUSION

Using coarse-grained molecular simulations of polyelectrolyte hydrogels and considering both short and long-range electrostatics, we demonstrate that a homogeneous hydrogel slab can undergo reversible size fluctuations when a time-dependent electric field is applied nonuniformly to a small part of the structure. The hydrogel can reduce its length by almost 50% with respect to its original size and retract until its size oscillation decays over time (Fig. 2). In simulations, we consider a semi-infinite hydrogel, where only one direction is finite, and the other two directions obey periodic boundary conditions such that there is an array of neighbouring slabs (Fig. 1). Such a system may correspond to an arrangement of hydrogel structures whose height is larger than its length. We anticipate that contraction can occur even for a single hydrogel since the distance separating two periodic images is as large as the hydrogel in our simulations. Notably, the contraction also occurs with weakly charged PE chains (i.e., smaller backbone charge fraction) and less polar solvents than we considered here (see SI text) but tend to diminish as the equilibrium state of the hydrogel becomes less swollen, for instance, by the addition of excess salt. Thus, we conclude that the reversible contraction phenomenon demonstrated here results from the interplay between stretched polyelectrolyte chains of hydrogel and the elasticity of these chains.

Our simulations also show that this contraction and retraction can occur quite rapidly for a wide range of frequencies and electric field strengths (i.e., on the scales of the relaxation time of the constituting polymer chains). While we only consider sinusoidal electric fields to prevent correlations between periodic simulation boxes, continuous signals can induce such contraction, possibly with a more oscillatory response. Notably, the frequency at which we observe the maximum contraction also coincides with the inverse of the relaxation time of the chains between two crosslinks. Thus, any signal faster than this time may not lead to a contraction response since chains cannot be stretched enough to deform the hydrogel.

To test if the phenomenon we observe is scalable, we repeat our simulations with a larger system, in which 100 unit cells are repeated in the $\hat{x}$ direction in the implicit solvent simulations. When the electric field is applied to a region with size $\lambda = L_0/10$, we observe the same contraction schemes, but the retraction times are longer as expected (see SI text).

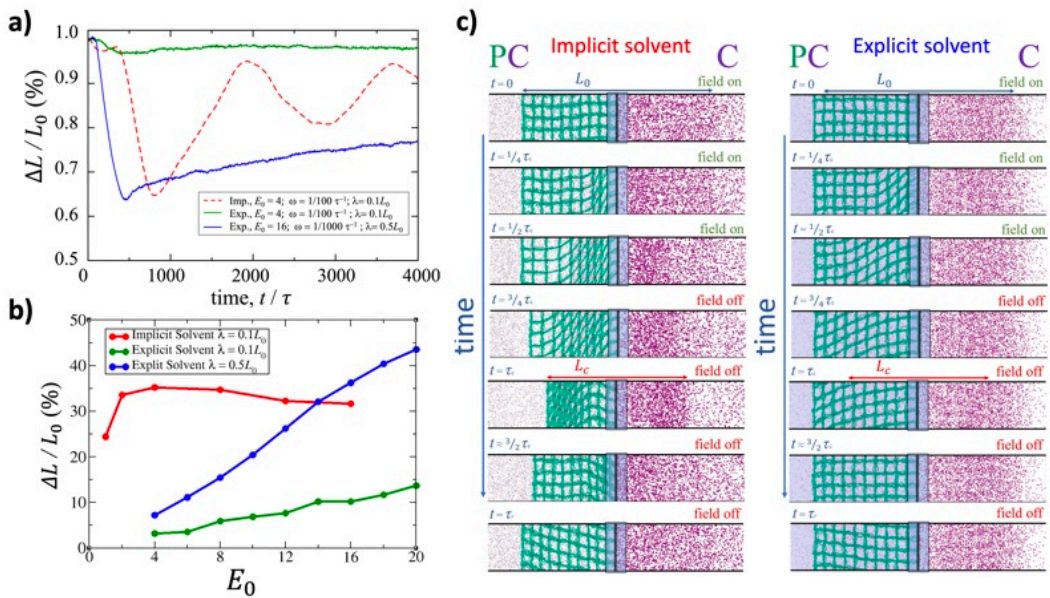

FIG. 5. Comparison between implicit and explicit solvent models. a) Time traces of rescaled hydrogel length as a function of simulation time for various frequency and field strengths. b) Recaled hydrogel length as a function of electric field strength for two solvent approaches. c) Representative simulation snapshots for implicit and explicit cases at consecutive time steps with identical frequencies and electric fields (i.e., $\omega = 100\tau^{-1}$ and $E_0 = 4\epsilon/(q\sigma)$) at $\lambda = 0.1L_0$. Notice that in the implicit case, there are no solvent beads.

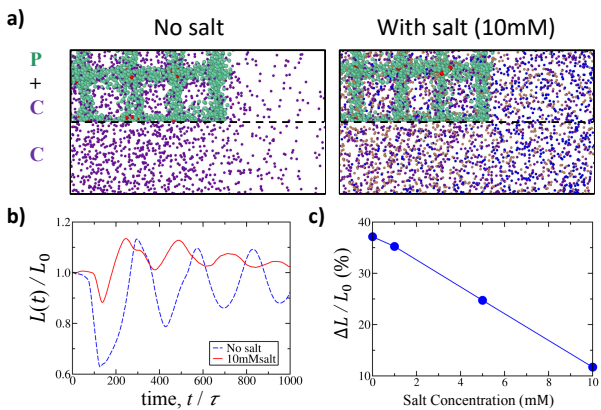

FIG. 6. Effect of excess salt on the hydrogel contraction in implicit solvent simulations. a) Representative simulation snapshots of a small section of the hydrogel slabs with no salt and 10 mM salt. Green and red beads indicate the PE backbone, purple beads indicate the counterions and blue and brown beads indicate the salt ions. b) Rescaled hydrogel size as a function of the rescaled time for two systems with and without salt. c) Contraction efficiency as a function of salt concentrations. The salt concentrations are given for the initial field-free conditions for consistency. $\omega = 100\tau^{-1}$ and $E_0 = \epsilon/(q\sigma)$ in all simulations.

Since our simulation results are given in reduced units, we may convert some of the numbers to practically more helpful real units: if we assume that Kuhn segments sizes of $b \approx 1$ nm and $b \approx 100$ nm for flexible and semiflexible PE chains, the length of our hydrogels $L_0 = 20 \times N \times b$ would be $L_0 \approx 200$ nm and $L_0 \approx 20$ $\mu$m, respectively. Increasing the polymerization degree, $N$, would result in even larger hydrogels. These two Kuhn segment sizes correspond to unit simulation times on the orders of $\tau \approx 1$ ns and $\tau \approx 1$ ms, which in turn would result in maximum contraction times of $\tau_c \approx 100$ ns and $\tau_c \approx 100$ ms, respectively, based on our intermediate frequency values given in Fig. 3. Notably, the retraction times increase with the dimensions of the hydrogel slab. Nevertheless, for several hundred micron-scale geometries, these time scales may be optimized to allow faster actuation metrics [30, 42].

As a final remark, one caveat in our simulations is that the dielectric constant is uniform throughout the simulations box. The presence of water or other polar solvents can alter the dielectric constant near the chains and also ionic distribution about individual chains [43, 44]. As a result, as the hydrogel deforms and relaxes, the ionic distribution could propagate across the hydrogel slab in a complex manner. Such time-dependent ionic response of hydrogels to mechanical distortions by explicitly including solvent molecules at the all-atom level can be the subject of our future studies.

### ACKNOWLEDGEMENTS

This research has been supported by TUBITAK 3501 Career program with grant number 119Z029.

# Supplementary Data and Figures for
# Controlling the shape of a Polyelectrolyte Hydrogel
# under Spatially Nonuniform Electric Fields

E. Mert Bahçeci[a] , Aykut Erbaş[b]

[a]Processing and Performance of Materials, Department of Mechanical Engineering, Eindhoven University of Technology, 5600 MB Eindhoven, The Netherlands

[b]UNAM-National Nanotechnology Research Center and Institute of Materials Science & Nanotechnology, Bilkent University, Ankara 06800, Turkey

# Simulation results of further work

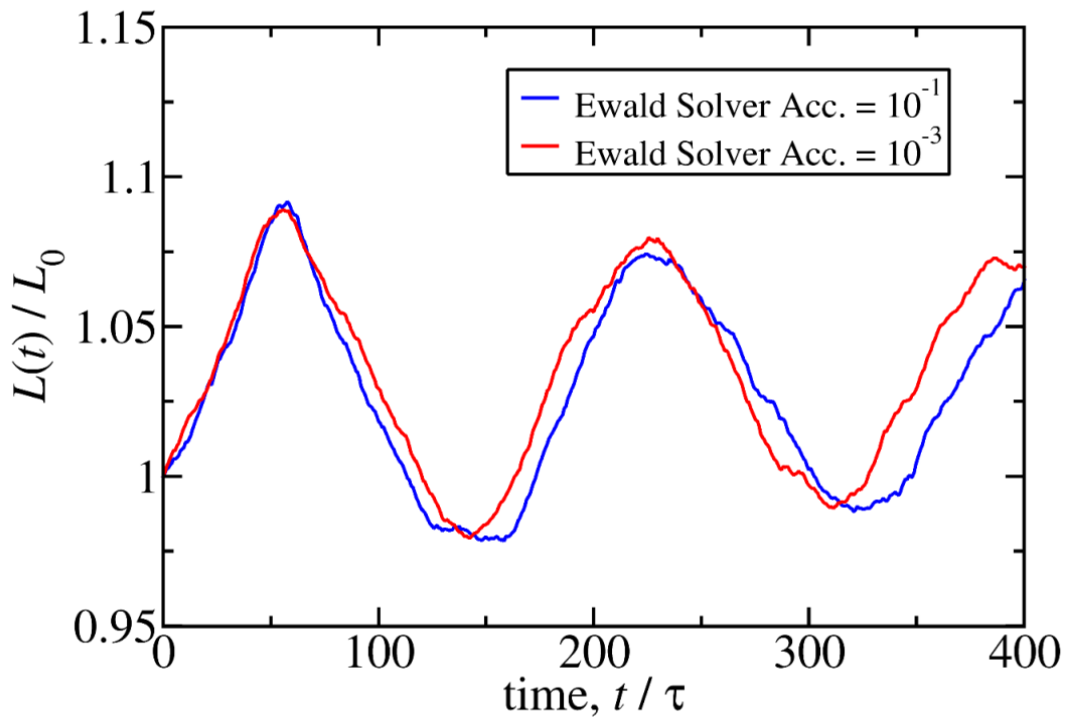

Figure 1: Electric field applied with amplitude of 0.5 and frequency of $1/100\tau$ to hydrogel system constructed as 20, 5, 3 units of lattices repeated in the $\hat{x}\ \hat{y}\ \hat{z}$ directions in implicit solvent simulations. With same relaxation conditions, different Ewald solver accuracies considered ($10^{-1}$ for blue, $10^{-3}$ for red line) to validate if low Ewald solver accuracy changes hydrogels behavior qualitatively.

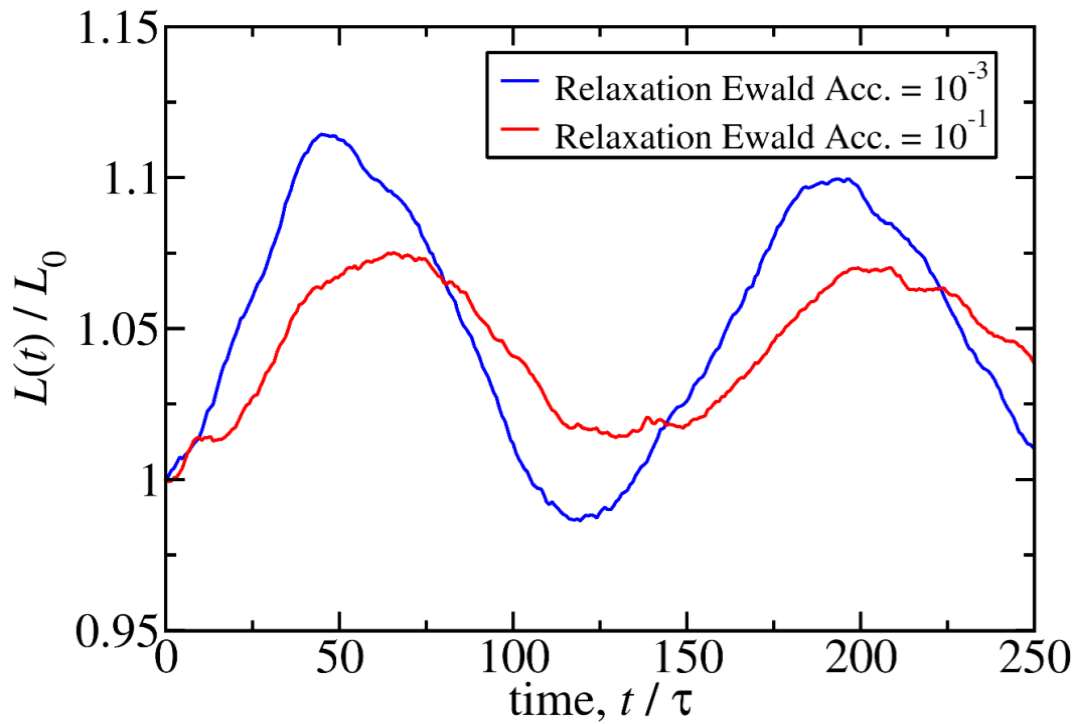

Figure 2: For hydrogel system constructed as 20, 5, 3 units of lattices repeated in the $\hat{x}\ \hat{y}\ \hat{z}$ directions, different Ewald solver accuracies compared in relaxation stage for different reduced LJ temperature ($T=1.0\ [\varepsilon/k_b]$) in implicit solvent simulations. Electric field applied after relaxation with 0.5 amplitude and $1/100\tau$ frequency. Qualitatively, there is no significant difference between different Ewald solver accuracies.

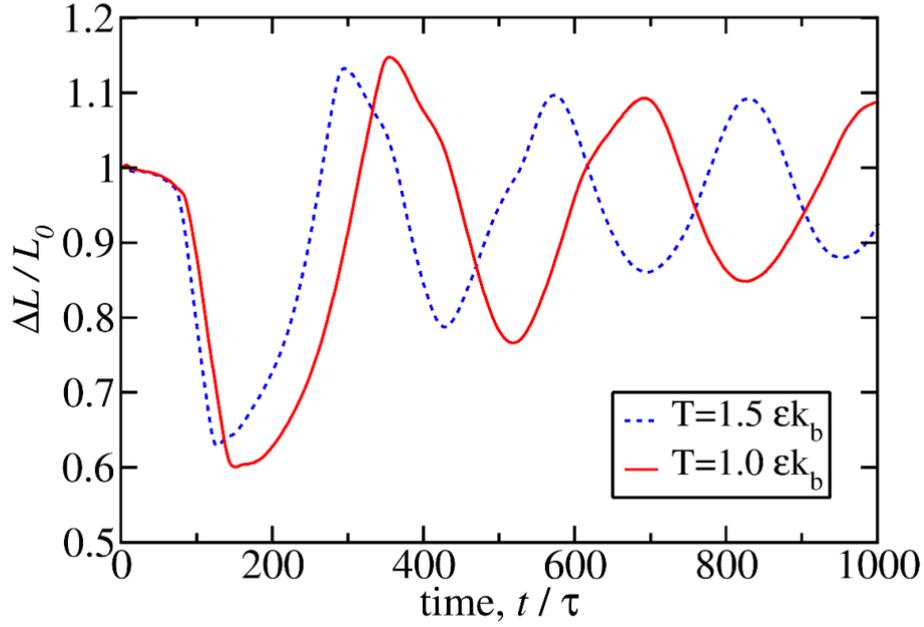

Figure 3: For the hydrogel system constructed as 40, 10, and 5 units of lattices repeated in the $\hat{x}$ $\hat{y}$ $\hat{z}$ directions, the effect of temperature was observed. The electric field applied after relaxation with 4.0 amplitude and $1/100\tau$ frequency. Qualitatively, there is no significant difference and quantitatively, shrinkage increased by 2.81% ($L_c$ 37.14% to 39.95% ) and contraction time increased by 19% ($\tau_c$ 128 to 152 [$t/\tau$]) when run at T=1.0 instead of T= 1.5 [$\varepsilon/k_b$]

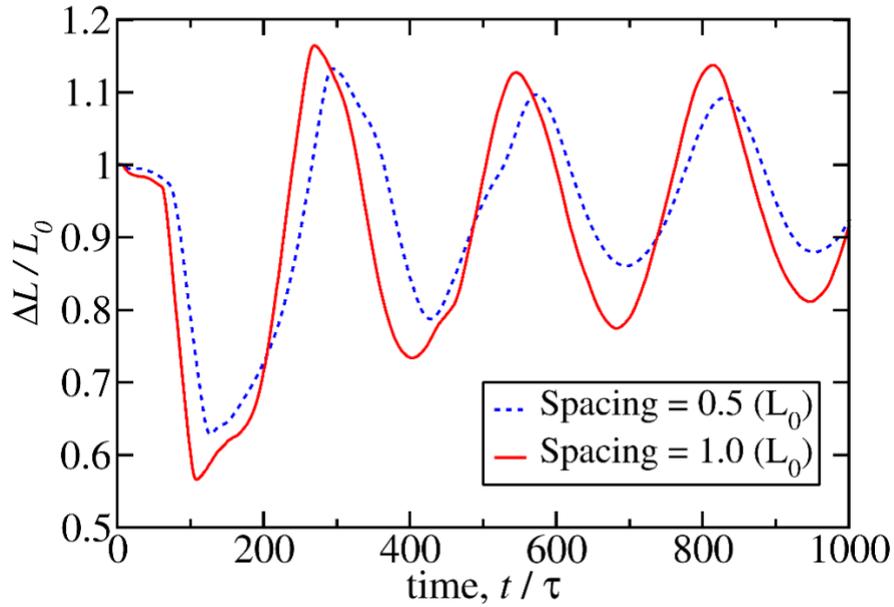

Figure 4: For hydrogel system constructed as 40, 10, 5 units of lattices repeated in the $\hat{x}$ $\hat{y}$ $\hat{z}$ directions in implicit solvent simulations, effect of different spacing between two hydrogels observed. Electric field applied after relaxation with 4.0 amplitude and $1/100\tau$ frequency. One with $0.5L_0$ and other with $1.0L_0$ between two hydrogels to see if electric field effect also have an impact to nearby hydrogels. Qualitatively, there is no significant difference and because this is the case with no salt, it should be also ok for systems with salt because there is higher electrostatic screening. Quantitatively, shrinkage increased by 6.30% ($L_c$ 37.14% to 43.44% ) and contraction time decreased by 14% ($\tau_c$ 128 to 108 [$t/\tau$]) when run at T=1.0 instead of T= 1.5 [$\varepsilon/k_b$]

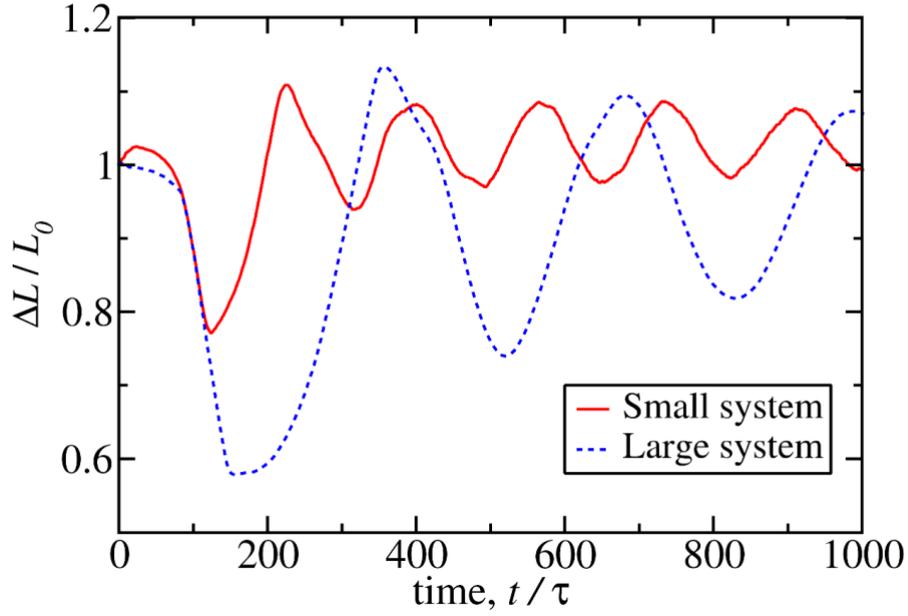

Figure 5: Comparison of small and large hydrogel systems, which constructed as 40, 10, 5 and 20, 5, 3 units of lattices repeated in the $\hat{x}$ $\hat{y}$ $\hat{z}$ directions, respectively, in implicit solvent simulations. Electric field applied after relaxation with 4.0 amplitude and $1/100\tau$ frequency where $T=1.0$ $[\varepsilon/k_b]$. Qualitatively, there is no significant difference and quantitatively, shrinkage decreased by 20% ($L_c$ 42.16% to 22.89%) and contraction time decreased by 21% ($\tau_c$ 156 to 123.2 $[t/\tau]$) when smaller system investigated instead of larger system.

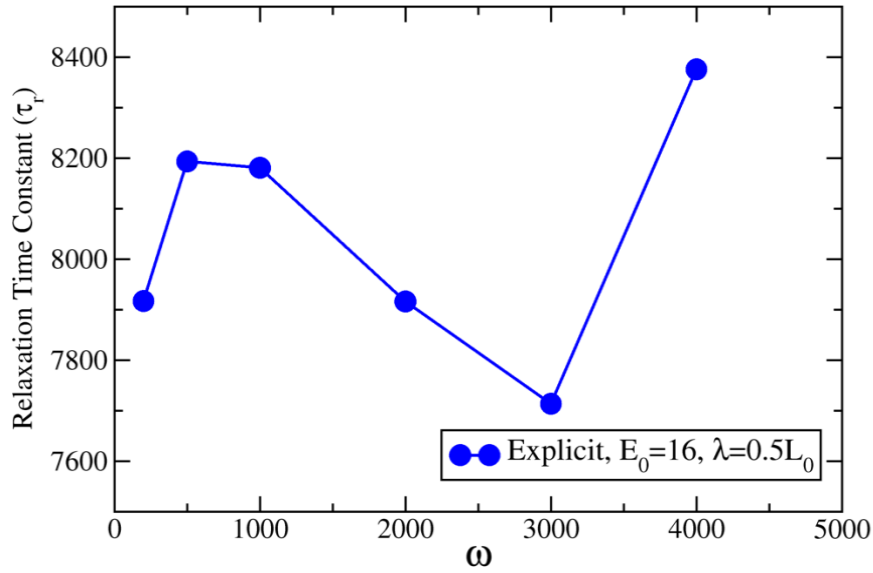

Figure 6: The retraction time of hydrogel in slab in explicit solvent simulations. The relaxation time is obtained via an exponential fit in the form of $f(x) = 1-A\exp(-t/B)$, here A and B are fit parameters.

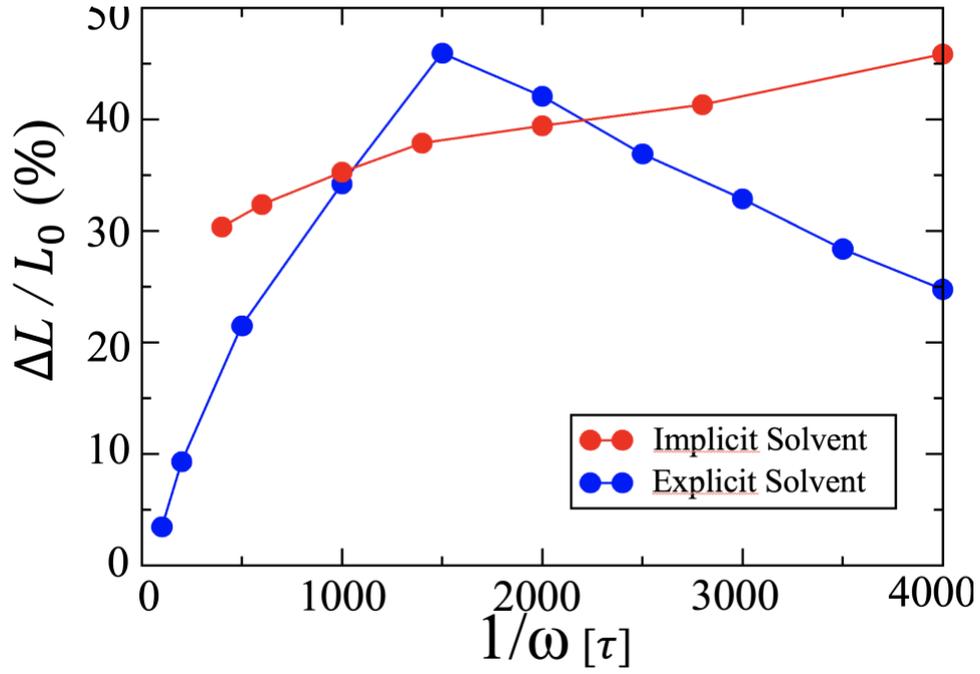

Figure 7: The contraction efficiency as a function of inverse frequency in explicit and implicit solvent simulations. E0 = 16 in reduced simulation units for all cases. However, the frequency axes for the implicit-simulation data is multiplied by 10 to provide a better visual presentation.

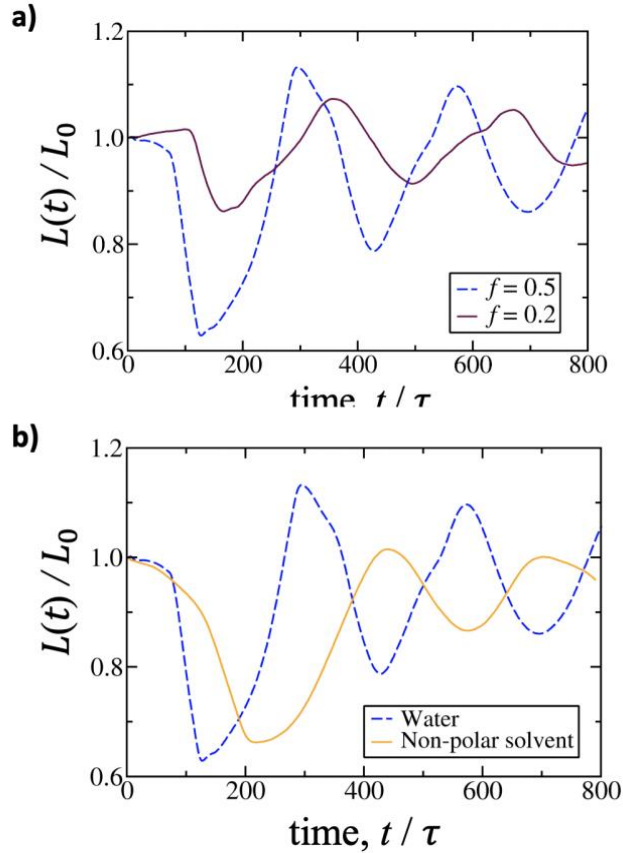

Figure 8: a) Contraction efficiency for two backbone charge fraction cases in implicit solvent simulations. The case with f=0.5 is used to obtain the data given in the main text. b) The contraction for two different dielectric cases. The case corresponding to water is used in the main text. In the non-polar case, the hydrogel is less swollen.

4